\documentclass[a4paper,dvips,12pt]{article}
\usepackage{epsfig,graphics,graphicx}
\newcommand{\rk}{\mbox{\boldmath $k$}}

\newcommand{\rp}{\mbox{\boldmath $p$}}

\begin{document}

\pagestyle{myheadings}
\vskip.5in
\begin{center}

%
%
\vskip.4in {\Large\bf Aspects of the Heavy-Quark Photoproduction in the Semihard Approach  }
\vskip.3in
%
%
%
C. Brenner Mariotto\footnote{Email: \tt mariotto@if.ufrgs.br}, M.B. Gay Ducati\footnote{Email: \tt gay@if.ufrgs.br }, M.V.T. Machado\footnote{Email: \tt magnus@if.ufrgs.br } \\
 Instituto de F\'{\i}sica, Universidade Federal do Rio Grande do
Sul. \\  Caixa Postal 15051, 91501-970 Porto Alegre, RS, BRAZIL.
%
%

%
\end{center}
%
\vskip.2in
\begin{abstract}
In this contribution we report on the calculations of   heavy-quark photoproduction using  the $k_{\perp}$-factorization (semihard) approach, emphasizing the results obtained with  the phenomenological saturation model.
\end{abstract}
%
%
%
\section{Introduction}

At high energies a new factorization theorem emerges, the $\rk_{\perp}$-factorization or semihard approach \cite{CCH,CE,GLRSS}. The relevant diagrams are considered with the virtualities and polarizations of the initial partons, taking into account the transverse momenta $q_{1\perp}$ and $q_{2\perp}$ of the incident partons. 
The processes are described through the convolution of off-shell matrix elements with the unintegrated parton distribution, ${\cal F}(x,\rk_{\perp})$. The latter can recover the usual parton distributions in the double logarithmic limit  by its integration over the transverse momentum $\rk_{\perp}$ exchanged gluon.   The matrix elements computed for the relevant subprocesses within this approach are more involved than those needed in the collinear approach already at LO level. On the other hand, a significant part of the NLO and some of the NNLO corrections to the LO contributions on the collinear approach, related to  the  contribution of non-zero transverse momenta of the incident partons, are already included in the LO contribution within the semihard approach \cite{RSS}. Moreover, part of the virtual corrections can be  resummed in the unintegrated gluon function \cite{RSS}. Furthermore, a very important issue is the consistency of the approach including nonleading-log effects and  the collinear factorization beyond leading order \cite{catani-hautmann}: 
the coefficient functions and the splitting functions providing $q(x,Q^2)$ and $G(x,Q^2)$ are supplemented with the all-order resummation of the $\alpha_s\ln (1/x)$ contributions at high energies, in contrast with a calculation in fixed order perturbation theory.

Two additional ingredients should be taken into account when calculating observables at high energies: treatment for the infrared sector and saturation effects. The unintegrated gluon function should evolve in transverse momentum through the BFKL evolution at high energies, leading to the diffusion on $\rk_{\perp}$ of the initial gluons in the evolution process. In this diffusion scenario 
the transverse momenta values are spread out into the infrared (and ultraviolet) region, where the perturbative description is not 
completely reliable.
The recently calculated non-linear corrections to the BFKL approach \cite{GB-Motyka-Stasto} introduce a natural treatment for these difficulties, where the saturation scale $Q_s$ provides the suitable cut-off controlling the infrared problems. As the Bjorken variable  $x\simeq Q^2/W^2$ decreases, unitarity corrections become important and 
control the steep growth of the gluon distribution. The most appealing approach taking into account both the notions of infrared behavior (confinement) and parton saturation phenomenon  is the saturation model \cite{GBW}, which is an eikonal-type model based on the color dipole picture of high energy interactions. In this contribution we discuss some aspects of the calculation of heavy quark photoproduction using the saturation model within the semihard approach. We address Ref. \cite{PRD66} for complete details and discussions. 

\section{Heavy-quark photoproduction in the $\rk_{\perp}$-factorization}

The differential cross section for the heavy-quark photoproduction process is expressed as the convolution of the unintegrated gluon function with the off-shell matrix elements  \cite{CCH,CataCiafaHaut,LiSaZot2000},
\begin{eqnarray}
  \frac{d\sigma (\gamma p \rightarrow Q\bar{Q}X)}{d^2 \rp_{1\perp}}   = \int dy_1^* \,d^2 \rk_{\perp} \frac{{\cal F}(x_2,\rk^2)\,\, |{\cal M}|^2(\mathrm{off\!-\!shell})}{\pi \alpha_2} \,,\label{eq:9}
  \end{eqnarray}
where the off-shell LO matrix elements are given by \cite{CataCiafaHaut,LiSaZot2000}.  The final expression for the photoproduction total cross section considering the direct component 
of the photon can be written as \cite{Shabel-Shuva},
\begin{eqnarray}
&\sigma_{tot}^{phot}& 
= 
  \frac{\alpha_{em}\,e_Q^2}{\pi}\, \int\, dz\,\,d^2 \rp_{1\perp} \, d^2\rk_{\perp} \, \frac{\alpha_s(\mu^2)\,{\cal F}(x_2,\rk_{\perp}^2; \mu^2)}{\rk_{\perp}^4}\nonumber \\
&&\!\!\!\!\!\!\times 
 \left\{ [z^2+ (1-z)^2]\,\left( \frac{\rp_{1\perp}}{D_1} + \frac{(\rk_{\perp}-\rp_{1\perp})}{D_2} \right)^2 +   m_Q^2 \,\left(\frac{1}{D_1} + \frac{1}{D_2}  \right)^2  \right\}\,, \label{eq:11} 
\end{eqnarray}
where $D_1 \equiv \rp_{1\perp}^2 + m_Q^2$ and 
$D_2 \equiv (\rk_{\perp}-\rp _{1\perp})^2 + m_Q^2$. Here,  $\alpha_{em}=1/137$ is the electromagnetic coupling constant and $e_Q$ is the electric charge of the produced heavy-quark. Details on the relevant variables and kinematics can be found at \cite{PRD66}. The scale $\mu$ in the strong coupling constant in general is taken to be equal to the gluon virtuality, $\mu^2=\rk^2$, in close connection with the BLM scheme \cite{BLM}. In the leading $\ln (1/x)$ approximation, $\alpha_s$ should take a constant value. When the transverse momenta of the incident partons are sufficiently smaller than those from the produced heavy-quarks, the result from the collinear approach is recovered. 

In Eq. (\ref{eq:11}) the unintegrated gluon function was allowed to depend also on the scale $\mu^2$, taken here as $\mu^2=\rp_{\perp}^2 + m_Q^2$, since some parametrizations take this scale into account in the computation of that quantity (see,  for instance Ref. \cite{lund-small-x}). The total and differential cross sections for the process can now be calculated, provided a suitable input for the function ${\cal F}(x_2,\rk_{\perp}^2; \mu^2)$. In what follows we use one of the  simplest parametrization available, covering a consistent treatment of the infrared region and taking into account the expected saturation effects  at high energies. These features are nicely  rendered in the phenomenological saturation model \cite{GBW}.  The unintegrated gluon distribution from that model, supplemented by the threshold factor (taking into account the large $x$ behavior),   is given by \cite{PRD66},
\begin{eqnarray}
 {\cal F}(x,\rk_{\perp}^2) = \frac{3\,\sigma_0}{4\,\pi^2 \alpha_s} \, R_0^2(x)\, \rk_{\perp}^4 \exp \left( -R_0^2(x)\,\rk_{\perp}^2  \right) \, (1-x)^7\,.
\label{eq:21}
\end{eqnarray}

\begin{figure}[t]
\centerline{\psfig{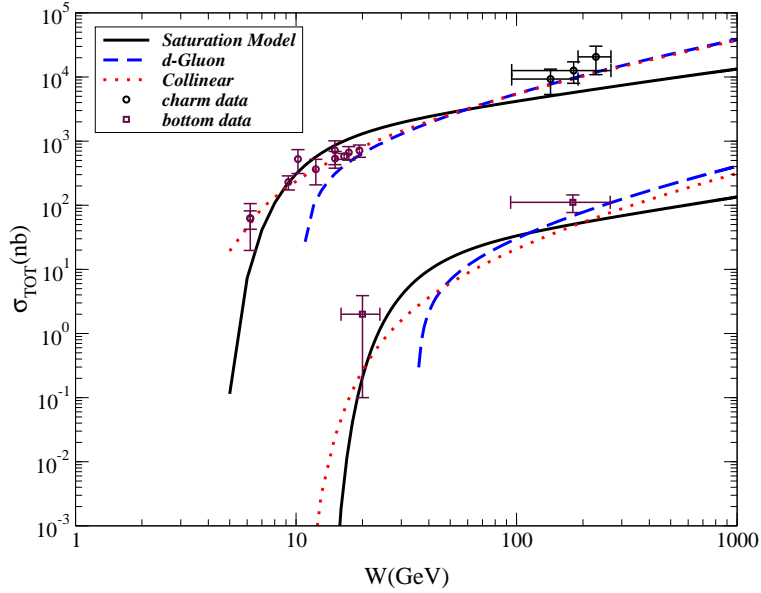}}
\caption{The results for the charm and bottom total cross sections considering the saturation model, the derivative of the collinear gluon distribution and the collinear parton model.}
\label{fig:sigtot} 
\end{figure}

We have used the parameters from \cite{GBW}, which includes the charm quark. In  Fig. (\ref{fig:sigtot}) the  charm and bottom total cross sections are presented. For sake of comparison, the saturation model is contrasted with the unintegrated gluon function given by the derivative of the collinear gluon distribution (labeled d-Gluon) \cite{PRD66}. The saturation model slightly underestimates high energy data, since the treatment of QCD evolution is not considered in the original model. Recent improvements, taking QCD evolution into account, should cure this shortcoming \cite{BGBK}. The derivative of the collinear gluon distribution gives a better description of high energy data, since it includes the referred gluon emission. The disagreement with the low energy data can be solved by introducing the  Sudakov form factor \cite{Martin}.   For sake of illustration, we also show the parton model results (collinear approach) for the LO process $\gamma g \rightarrow Q\bar{Q}$, where it has been used $m_c=1.3$ GeV, $m_b=4.75$ GeV, and $\mu^2=\hat{s}$.
This gives a reasonable description of data given the use of lower heavy quark masses or alternatively considering higher order corrections to the LO calculation.
In contrast, the semihard approach gives a reasonable 
description of data already at LO level. 
The energy dependence is distinct in the calculations: the saturation model provides a mild energy growth, whereas in the collinear approach the growth is steeper.  A related study considering also the resolved photon component and extended calculations to the heavy quark production in two-photon process can be found in Ref. \cite{MotykaTimneanu}.
An analysis of the heavy quark transverse momentum distributions  can also be found in Ref. \cite{PRD66}.

\begin{thebibliography}{99}

\bibitem{CCH}
S.~Catani, M.~Ciafaloni, F.~Hautmann,  Nucl. Phys. {\bf B366}, 135 (1991).

\bibitem{CE}
J.~Collins, R.~Ellis, Nucl. Phys. {\bf B360}, 3 (1991).

\bibitem{GLRSS}
L.~Gribov, E.~Levin, M.~Ryskin,  Phys. Rep. {\bf 100}, 1 (1983); E.M. Levin, M.G. Ryskin, Y.M. Shabelski, A.G. Shuvaev, Sov. J. Nucl.
  Phys. {\bf 53}, 657 (1991).


\bibitem{RSS} M.G.~Ryskin, A.G.~Shuvaev, Y.M.~Shabelski,
Phys.\ Atom.\ Nucl.\  {\bf 64}, 1995 (2001); 

\bibitem{catani-hautmann}
S.~Catani and F.~Hautmann, Nucl. Phys. B {\bf 427}, 475 (1994).


\bibitem{GB-Motyka-Stasto}
K.~Golec-Biernat, L.~Motyka and A.M.~Stasto, 
Phys.\ Rev.\ D {\bf 65}, 074037 (2002).

\bibitem{GBW}
K.~Golec-Biernat, M.~W\"usthoff,  Phys. Rev.  D {\bf 59}, 014017 (1998).

\bibitem{PRD66} C. Brenner Mariotto, M.B. Gay Ducati, M.V.T. Machado, Phys. Rev. {\bf D} 66, 114013 (2002). 

\bibitem{CataCiafaHaut} S.~Catani, M.~Ciafaloni and F.~Hautmann, Phys.\ Lett.\ B {\bf 242}, 97 (1990).

\bibitem{LiSaZot2000}
A.V. Lipatov, V.A.~Saleev, N.P. Zotov,  Mod. Phys. Lett. A {\bf 15}, 1727 (2000).


\bibitem{Shabel-Shuva}
Y.M.~Shabelski, A.G.~Shuvaev, Los Alamos preprint [hep-ph/0107106].


\bibitem{BLM} S.J.Brodsky, G.P. Lepage, P.B.Mackenzie,  Phys. Rev. D {\bf 28}, 228 (1983).

\bibitem{lund-small-x}
B.~Andersson {\it et al.}  [Small $x$ Collaboration], Eur.\ Phys.\ J.\ C 
 {\bf 25}, 77 (2002).

\bibitem{BGBK}
J.~Bartels, K.~Golec-Biernat, H.~Kowalski,
Phys.\ Rev.\ D {\bf 66}, 014001 (2002).

\bibitem{Martin} M.A. Kimber, A.D. Martin, M.G. Ryskin, Phys. Rev. {\bf D} 63, 114027 (2001).

\bibitem{MotykaTimneanu} L. Motyka, N. Timneanu,  Eur.\ Phys.\ J.\ C ({\it in press}), [hep-ph/0209029].

\end{thebibliography}
\end{document}